\documentclass[cits]{PoS}
\usepackage{amsmath}

\newcommand{\al}{\ensuremath{\alpha} }
\newcommand{\be}{\ensuremath{\beta} }
\newcommand{\ga}{\ensuremath{\gamma} }
\newcommand{\De}{\ensuremath{\Delta} }
\newcommand{\ka}{\ensuremath{\kappa} }
\newcommand{\la}{\ensuremath{\lambda} }
\newcommand{\lalat}{\ensuremath{\la_{\text{lat}}} }
\newcommand{\si}{\ensuremath{\sigma} }
\newcommand{\cD}{\ensuremath{\mathcal D} }
\newcommand{\cN}{\ensuremath{\mathcal N} }
\newcommand{\cO}{\ensuremath{\mathcal O} }

\newcommand{\cQ}{\ensuremath{\mathcal Q} }
\newcommand{\Rbb}{\ensuremath{\mathbb R} }
\newcommand{\cU}{\ensuremath{\mathcal U} }
\newcommand{\Gdag}{\ensuremath{G^{\dag}} }
\newcommand{\Qbar}{\ensuremath{\overline Q} }
\newcommand{\cUbar}{\ensuremath{\overline{\mathcal U}} }
\newcommand{\aldot}{\ensuremath{\dot\al} }
\newcommand{\muhat}{\ensuremath{\widehat\mu} }
\newcommand{\SO}[1]{\ensuremath{\text{SO(}#1\text{)}} }
\newcommand{\SU}[1]{\ensuremath{\text{SU(}#1\text{)}} }
\newcommand{\U}[1]{\ensuremath{\text{U(}#1\text{)}} }
\newcommand{\Uone}{\ensuremath{\text{U(1)}} }
\newcommand{\vev}[1]{\ensuremath{\left\langle #1 \right\rangle} }
\newcommand{\gsim}{\ensuremath{\gtrsim} }
\newcommand{\lsim}{\ensuremath{\lesssim} }
\newcommand{\llra}{\ensuremath{\longleftrightarrow} }
\newcommand{\X}{\ensuremath{\!\times\!} }
\newcommand{\pf}[0]{\ensuremath{\mbox{pf}\,} }
\newcommand{\Tr}[1]{\ensuremath{\mbox{Tr}\left[ #1 \right]} }

\newcommand{\fig}[1]{Fig.~\ref{#1}}
\newcommand{\secref}[1]{Sec.~\ref{#1}}
\newcommand{\refcite}[1]{Ref.~\cite{#1}}

\title{Progress and prospects of lattice supersymmetry}
\ShortTitle{Progress and prospects of lattice supersymmetry}
\author{\speaker{David Schaich} \\
  AEC Institute for Theoretical Physics, University of Bern, 3012 Bern, Switzerland \\
  E-mail: \email{schaich@itp.unibe.ch}
}

\abstract{ 
  Supersymmetry plays prominent roles in the study of quantum field theory and in many \mbox{proposals} for potential new physics beyond the standard model, while lattice field theory provides a non-perturbative regularization suitable for strongly interacting systems.
  Lattice investigations of supersymmetric field theories are currently making significant progress, though many challenges remain to be overcome.
  In this brief overview I discuss particularly notable progress in three \mbox{areas:} supersymmetric Yang--Mills (SYM) theories in fewer than four dimensions, as well as both minimal $\cN = 1$ SYM and maximal $\cN = 4$ SYM in four dimensions.
  I also highlight super-QCD and sign problems as prominent challenges that will be important to address in future work.
}

\FullConference{36th Annual International Symposium on Lattice Field Theory \\ 22--28 July 2018 \\ Michigan State University, United States}

\begin{document}
\setlength{\abovedisplayskip}{6 pt}
\setlength{\belowdisplayskip}{6 pt}
\setcounter{section}{-1}
\section{\label{sec:intro}Introduction, motivation and background} 
Supersymmetry plays prominent roles in modern theoretical physics, as a tool to improve our understanding of quantum field theory (QFT), as an ingredient in many new physics models, and as a means to study quantum gravity via holographic duality.
Lattice field theory provides a non-perturbative regularization for QFTs, and other contributions to these proceedings document the prodigious success of this framework applied to QCD and similar theories. 
It is natural to consider employing lattice field theory to investigate supersymmetric QFTs, especially in strongly coupled regimes.
In this proceedings I review the recent progress and future prospects of lattice studies of supersymmetric systems, focusing on four-dimensional gauge theories and their dimensional reductions to $d < 4$.\footnote{Theories without gauge invariance, such as Wess--Zumino models and sigma models, are reviewed in Refs.~\cite{Catterall:2009it, Kadoh:2016eju}.  More recent work in this area includes Refs.~\cite{Aoki:2017iwi, Kadoh:2018hqq, Kadoh:2018ele, Kadoh:2018ivg}.} 

Lattice supersymmetry now has more than four decades of history~\cite{Dondi:1976tx}, much of which is reviewed by Refs.~\cite{Catterall:2009it, Giedt:2009yd, Joseph:2015xwa, Kadoh:2016eju, Bergner:2016sbv, Hanada:2016jok}.
Unfortunately, progress in this field has been slower than for QCD, in large part because the lattice discretization of space-time breaks supersymmetry.
This occurs in three main ways.
First, the anti-commutation relation $\left\{Q_{\al}, \Qbar_{\aldot}\right\} = 2\si_{\al\aldot}^{\mu} P_{\mu}$ in the super-Poincar\'e algebra connects the spinorial generators of supersymmetry transformations, $Q_{\al}$ and $\Qbar_{\aldot}$, to the generator of infinitesimal space-time translations, $P_{\mu}$.
The absence of infinitesimal translations on the lattice consequently implies broken supersymmetry.

Next, bosonic and fermionic fields are typically discretized differently on the lattice (in part due to the famous fermion doubling problem). 
In the specific context of supersymmetric gauge theories, standard discretizations associate the gauginos with lattice sites $n$ (i.e., they transform as $G(n) \la_{\al}(n) \Gdag(n)$ under a lattice gauge transformation) while the gauge connections are associated with links between nearest-neighbor sites, transforming as $G(n) U_{\mu}(n) \Gdag(n + a\muhat)$ where `$a$' is the lattice spacing.
Away from the $a \to 0$ continuum limit, these differences prevent supersymmetry transformations from correctly interchanging superpartners.

Finally, supersymmetry requires a derivative operator that obeys the Leibniz rule~\cite{Dondi:1976tx}, viz.\ $\partial \left[\phi \eta\right] = \left[\partial \phi\right]\eta + \phi \partial \eta$, which is violated by standard lattice finite-difference operators.
`No-go theorems' presented by Refs.~\cite{Kato:2008sp, Bergner:2009vg} establish that only non-local derivative and product operators can obey the Leibniz rule (and hence fully preserve supersymmetry) in discrete space-time.
Efforts continue to construct and study alternate formulations that may better balance locality and supersymmetry.
For example, \refcite{Kato:2016fpg} finds that a lattice field product operator obeying a `cyclic Leibniz rule'~\cite{Kato:2013sba} suffices to preserve partial supersymmetry and establish non-renormalization for a quantum-mechanical system. 
\refcite{DAdda:2017bzo} introduces a different non-local `star product' that preserves the Leibniz rule, with the consequence that the lattice spacing no longer acts as a regulator. 
Despite the complicated intricacies of these constructions, so far their applicability appears limited to systems without gauge invariance, and only in (0+1)~dimensions~\cite{Kato:2016fpg, Kato:2018kop} or on infinite lattices~\cite{DAdda:2017bzo}.

As a consequence of broken supersymmetry, quantum effects in the lattice calculation will generate supersymmetry-violating operators.
These include, in particular, relevant operators for which counterterms will have to be fine-tuned in order to recover the supersymmetric QFT of interest in the $a \to 0$ continuum limit that corresponds to removing the UV cutoff $a^{-1}$.
In theories with scalar fields (either squarks in super-QCD or scalar elements of the gauge supermultiplet in $\cN > 1$ theories with extended supersymmetry), these scalars' mass terms present fine-tuning problems similar to that of the Higgs boson in the standard model.
Additional supersymmetry-violating operators include fermion (quark and gaugino) mass terms, Yukawa couplings, and quartic (four-scalar) terms.
Altogether there are typically $\cO(10)$ of these operators~\cite{Giedt:2009yd, Elliott:2008jp, Catterall:2014mha}, implying such high-dimensional parameter spaces that there seems to be little hope of effectively navigating them in numerical lattice calculations.

The following three sections focus on three different ways to reduce the amount of fine-tuning in lattice studies of supersymmetric Yang--Mills (SYM) theories.
We begin in the next section by reviewing dimensional reductions to SYM theories in fewer than four space-time dimensions, which has received the most attention from the community so far.
We return to four dimensions in \secref{sec:min}, considering first the special case of minimal ($\cN = 1$) SYM, which is vastly simplified by the absence of scalar fields.
Another special case in four dimensions is maximal ($\cN = 4$) SYM, the topic of \secref{sec:max}, for which a closed subalgebra of the supersymmetries can be preserved at non-zero lattice spacing, again drastically reducing the necessary fine-tuning.
Finally, \secref{sec:future} briefly discusses some prominent challenges to be faced by lattice studies of supersymmetric QFTs in the future, including investigations of supersymmetric QCD (SQCD) and the possibility of sign problems in various theories.

\section{\label{sec:lowd}Lower-dimensional systems} 
Dimensionally reduced SYM theories can be much easier to analyze numerically. 
In addition to the smaller number of degrees of freedom for an $L^d$ lattice, the resulting lower-dimensional theories tend to be super-renormalizable and in many cases a one-loop counterterm calculation suffices to restore supersymmetry in the continuum limit~\cite{Giedt:2004vb, Bergner:2007pu, Giedt:2018ygt}. 
We will label systems by their number of supercharges (generators of supersymmetry transformations): $Q = 4$, 8 or 16 respectively corresponding to $\cN = 1$, 2 or 4 SYM in four dimensions (or equivalently to minimal SYM in $d = 4$, 6 or 10 dimensions).
For $d \leq 4$ these theories involve a gauge field, $Q$ fermionic component fields, and $4 - d$, $6 - d$ or $10 - d$ real scalar fields, respectively, all of which are massless and transform in the adjoint representation of the gauge group.
The gauge groups we consider are SU($N$) and $\U{N} = \SU{N} \otimes \Uone$.

\paragraph{0+1 dimensions:} 
The reduction to `SYM quantum mechanics' (QM) has been the subject of many numerical studies over the past decade, starting with Refs.~\cite{Hanada:2007ti, Catterall:2007fp}. 
These systems involve balanced collections of interacting bosonic and fermionic $N\X N$ matrices at a single spatial point.
One reflection of the simplicity of SYM QM is that a lattice regularization may not even be required; a gauge-fixed Monte Carlo approach employing a hard momentum cutoff~\cite{Hanada:2007ti} was used by Refs.~\cite{Anagnostopoulos:2007fw, Hanada:2008gy, Hanada:2008ez, Hanada:2009ne, Hanada:2011fq, Hanada:2013rga}. 
Another illustration is a recent proposal~\cite{Maldacena:2018vsr} that `ungauging' $Q = 16$ SYM QM (to consider a scalar--fermion system with SU($N$) global symmetry) has relatively little effect, in the sense that both the gauged and ungauged models flow to the same theory in the IR.
This conjecture was quickly tested by lattice calculations that found consistent results~\cite{Berkowitz:2018qhn}.

\begin{figure}[bp]
  \centering
  \includegraphics[height=0.36\linewidth]{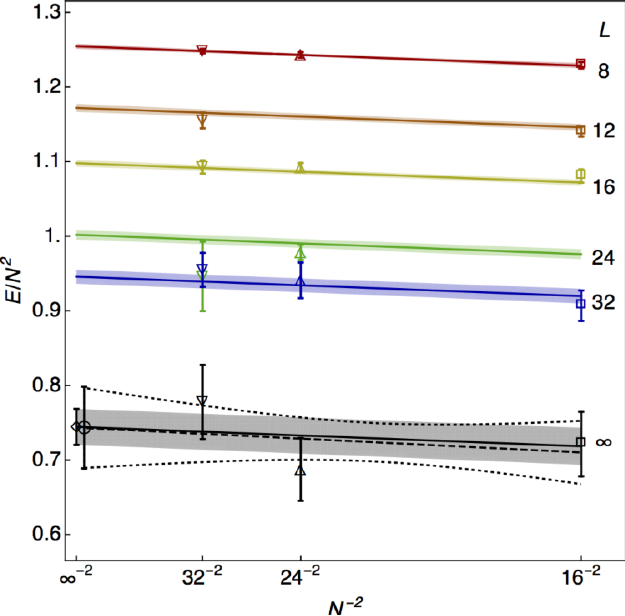}\hfill \includegraphics[height=0.36\linewidth]{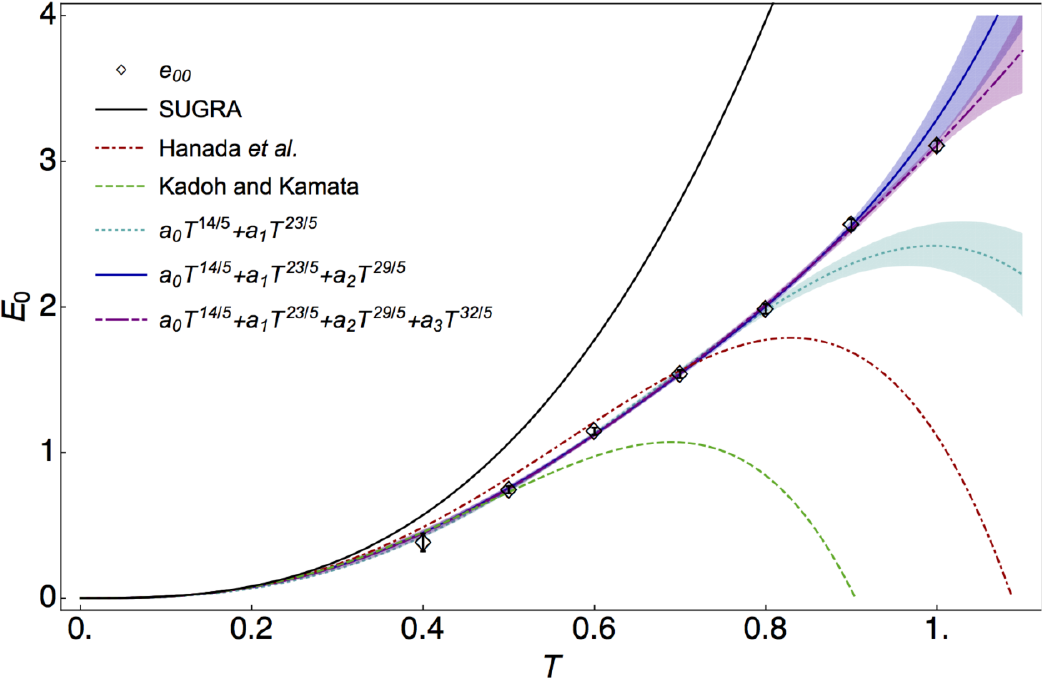}
  \caption{\label{fig:BFSS}State-of-the-art results for the dual black hole internal energy from $Q = 16$ SYM QM lattice calculations, from Refs.~\protect\cite{Berkowitz:2016tyy, Berkowitz:2016jlq}.  \textbf{Left:} Representative (separate and combined) large-$N$ and continuum ($L \to \infty$) extrapolations, for fixed dimensionless temperature $T = 0.5$.  \textbf{Right:} Large-$N$ continuum-limit results versus $T$, compared to earlier investigations~\protect\cite{Hanada:2008ez, Kadoh:2015mka} that did not carry out controlled extrapolations.} 
\end{figure}

Even though these quantum-mechanical systems are much simpler to study on the lattice than their $d = 4$ SYM counterparts, they remain computationally non-trivial.
This is demonstrated by the state-of-the-art results for $Q = 16$ SYM QM from Refs.~\cite{Berkowitz:2016tyy, Berkowitz:2016jlq} shown in \fig{fig:BFSS}.
This $Q = 16$ case has attracted particular interest due to its connections to string theory~\cite{deWit:1988wri}, and especially the conjecture~\cite{Banks:1996vh} that the large-$N$ limit of this system describes the strong-coupling (`M-theory') limit of type-IIA string theory in light-front coordinates. 
(Refs.~\cite{Taylor:1999qk, Ydri:2017ncg} are thorough reviews of the stringy side.) 
At finite temperature, this conjecture relates the large-$N$ limit of the (deconfined) $Q = 16$ system to a dual compactified 11-dimensional black hole geometry in M-theory, and \fig{fig:BFSS} shows the dual black hole internal energy determined from lattice SYM QM computations.
This quantity was previously investigated numerically by Refs.~\cite{Anagnostopoulos:2007fw, Hanada:2008ez, Hanada:2013rga, Catterall:2008yz, Catterall:2009xn, Kadoh:2015mka, Filev:2015hia, Hanada:2016zxj}. 

Refs.~\cite{Berkowitz:2016tyy, Berkowitz:2016jlq} improve upon the earlier work by carrying out controlled extrapolations to the large-$N$ continuum limit, allowing for more robust comparisons with dual gravitational predictions.
The left plot of \fig{fig:BFSS} shows one such extrapolation, for a fixed value $T = 0.5$ of the dimensionless temperature $T \equiv t_{\text{dim}} / \la_{\text{dim}}^{1 / 3}$.
(The subscripts highlight dimensionful quantities, including the 't~Hooft coupling $\la_{\text{dim}} = g^2 N$ with dimension $[\la_{\text{dim}}] = 4 - d$.)
With fixed $T$ the continuum limit corresponds to extrapolating the number of lattice sites $L \to \infty$.
At low temperatures the results in the right plot convincingly approach the leading-order gravitational prediction from classical supergravity (SUGRA), providing non-perturbative first-principles evidence that the holographic duality conjecture is correct.
In addition, the growing difference between the lattice results and the SUGRA curve at higher temperatures can be considered a prediction of higher-order quantum gravitational effects that are enormously difficult to calculate analytically.

The computational non-triviality of these investigations comes primarily from the large values of $N$ that are needed ($16 \leq N \leq 32$ in Refs.~\cite{Berkowitz:2016tyy, Berkowitz:2016jlq}, large enough to benefit from dividing individual $N\X N$ matrices across multiple MPI processes via the \texttt{\href{https://sites.google.com/site/hanadamasanori/home/mmmm}{MMMM}} code).
The computational cost of $N\X N$ matrix multiplication scales $\propto$$N^3$, compared to the $\sim$$L^{5d / 4}$ costs of the rational hybrid Monte Carlo (RHMC) algorithm.
In addition to improving control over the $N \to \infty$ extrapolations, large values of $N \gsim 10$ are also required to suppress a thermal instability associated with the non-compact quantum moduli space of $Q = 16$ SYM QM~\cite{Catterall:2009xn}.
For sufficiently low temperatures and sufficiently small $N$ the system is able to run away along these flat directions (holographically interpreted as D0-brane radiation from the dual black hole).
Formally a scalar potential should be added to the lattice action to stabilize the desired vacuum, and then removed in the course of the continuum extrapolation, further increasing computational costs~\cite{Catterall:2009xn, Hanada:2009hq}.
However, Refs.~\cite{Berkowitz:2016tyy, Berkowitz:2016jlq} argue that in practice it is possible to carry out Monte Carlo sampling around a metastable vacuum so long as $N$ is sufficiently large.
In particular, $N$ must increase in order to reach smaller $T$.

Further numerical investigations of SYM QM systems are underway~\cite{Steinhauer:2014oda, Ambrozinski:2014oka, Bergner:2016qbz, Rinaldi:2017mjl, Buividovich:2018scl}. 
At the same time, the good control over the necessary extrapolations that has now been achieved for the $Q = 16$ case also motivates pursuing comparable quality in lattice studies of less-simplified systems.
One example of such a system is the Berenstein--Maldacena--Nastase (BMN) deformation of $Q = 16$ SYM QM~\cite{Berenstein:2002jq}, which introduces a non-zero mass for the 9 scalars and 16 fermions while preserving all 16 supercharges. 
This theory has been studied numerically by Refs.~\cite{Catterall:2010gf, Honda:2013nfa, Asano:2018nol}. 
The mass deformation explicitly breaks the SO(9) global symmetry (corresponding to the compactified spatial dimensions of $d = 10$ SYM) down to $\SO{6}\X \SO{3}$.
It also lifts the flat directions mentioned above, thus serving as a supersymmetric regulator that need not be removed in the continuum limit.

\begin{figure}[bp]
  \centering
  \includegraphics[height=0.32\linewidth]{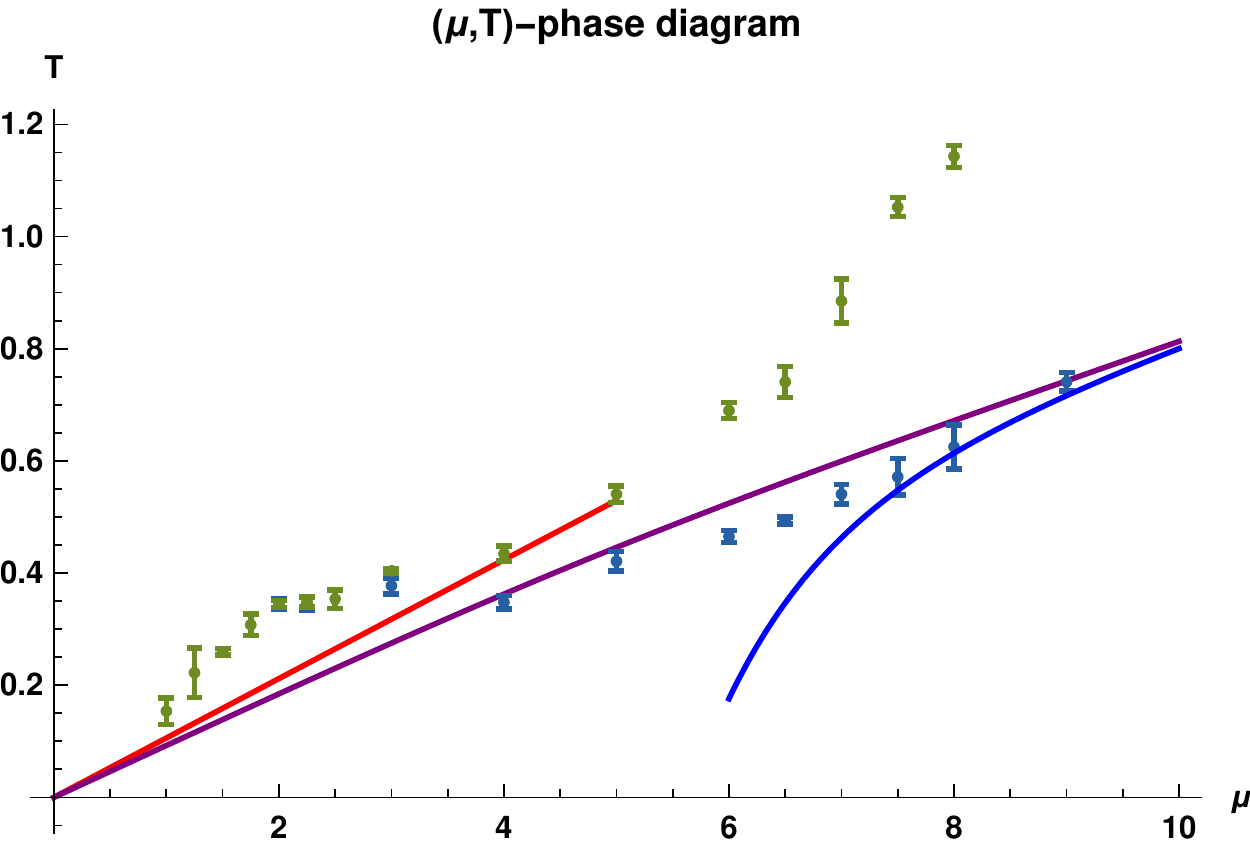}\hfill \includegraphics[height=0.32\linewidth]{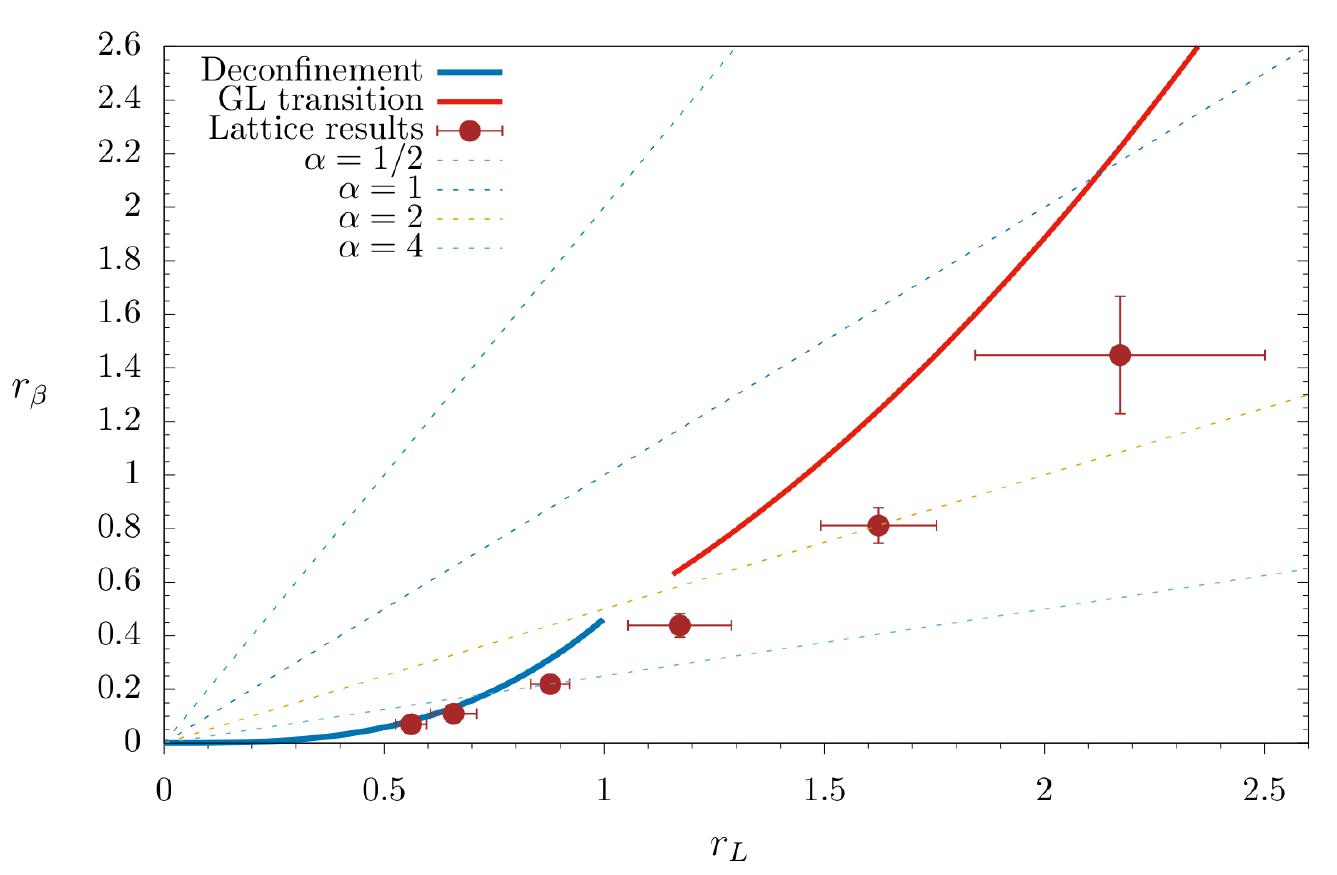}
  \caption{\label{fig:phase_diagrams}\textbf{Left:} Phase diagram for the BMN deformation of $Q = 16$ SYM QM, in the plane of dimensionless mass $\mu \equiv \mu_{\text{dim}} / \la_{\text{dim}}^{1 / 3}$ and temperature $T$, from \protect\refcite{Asano:2018nol}.  Lattice results for the confinement (lower blue points) and $\SO{9} \to \SO{6}\X \SO{3}$ (upper green points) transitions with fixed $N = 8$ and $L = 24$ are compared to a small-$\mu$ holographic calculation (red), three-loop large-$\mu$ perturbation theory (blue) and an interpolating resummation (purple).  \textbf{Right:} Phase diagram for two-dimensional $Q = 16$ SYM on an $r_L \X r_{\be} = r_L \X \frac{1}{T}$ torus, from \protect\refcite{Catterall:2017lub}.  Lattice results for the `spatial deconfinement' transition with fixed $N = 12$ and aspect ratios $\al = r_L / r_{\be}$ from 8 ($32\X 4$) to $3 / 2$ ($18\X 12$) are compared to the large-$N$ high-$T$ bosonic QM behavior (blue) and a low-$T$ holographic calculation (red).}
\end{figure}

In addition, the mass parameter $\mu$ provides a second axis for the finite-temperature phase diagram, as shown in the left plot of \fig{fig:phase_diagrams}.
As $\mu \to \infty$ the theory becomes gaussian, and the deconfinement temperature $T_d$ can be computed perturbatively in $1 / \mu$.
At small $\mu$, \refcite{Costa:2014wya} carried out the numerical construction of the SUGRA black hole geometry dual to the deconfined phase, predicting $T_d$ to linear order in $\mu$.
Figure~\ref{fig:phase_diagrams} shows recent numerical results from \refcite{Asano:2018nol} in reasonably good agreement with these predictions, given the fixed $N = 8$ and $L = 24$.
In addition to the deconfinement transition signalled by the Polyakov loop, this work observes a transition between an approximately SO(9)-symmetric phase at high temperatures and an $\SO{6}\X \SO{3}$ phase at low temperatures.
For small $\mu \lsim 3$ these transitions occur at the same $T_d$, while at larger $\mu$ higher temperatures are needed to recover approximate SO(9) symmetry.
It will be interesting to systematize large-$N$ continuum extrapolations in future lattice BMN investigations, since these turned out to be significant in the $\mu = 0$ limit considered in \fig{fig:BFSS}.

\paragraph{1+1 dimensions:} 
Dimensional reductions of SYM to $d = 2$ and 3 also provide less-simplified systems compared to SYM QM, while still being significantly more tractable than $d = 4$.
Although there has been a lot of work in this area over the years, much of the effort has focused on constructing clever lattice formulations that minimize fine-tuning in principle, rather than using these constructions in practical numerical calculations.
Here we will highlight the numerical calculations, which leaves little to say about $d = 3$: see Refs.~\cite{Giedt:2018ygt, Catterall:2011cea, Giedt:2017fck} for $Q = 8$ formulations. 

The main clever constructions that have been applied are based on either `twisting'~\cite{Sugino:2003yb, Sugino:2004qd, Catterall:2004np} or orbifolding~\cite{Cohen:2003xe, Cohen:2003qw, Kaplan:2005ta, Unsal:2006qp}, two approaches that actually produce equivalent constructions~\cite{Catterall:2007kn, Damgaard:2008pa}.
(See \refcite{Catterall:2009it} for a thorough review.)
Here we discuss only the twisting approach, which identifies at most $\lfloor Q / 2^d \rfloor$ linear combinations of supercharges, $\cQ$, that are nilpotent, $\cQ^2 = 0$.
These are found by organizing the $Q$ supercharges into irreducible representations of a `twisted rotation group' $\SO{d}_{\text{tw}} \equiv \mbox{diag}\left[\SO{d}_{\textrm{euc}} \otimes \SO{d}_R\right]$, where $\SO{d}_{\textrm{euc}}$ is the Wick-rotated Lorentz group and $\SO{d}_R$ is a global $R$-symmetry.
The nilpotent \cQ are those that transform in the twisted-scalar representation.
The requirement $Q \geq 2^d$ ensures a sufficiently large $R$-symmetry.
This procedure provides a closed supersymmetry subalgebra $\left\{\cQ, \cQ\right\} = 0$ at non-zero lattice spacing, leading to a $\cQ$-invariant lattice action with no need of the Leibniz rule.

For some theories there are multiple ways the twisting procedure can be carried out.
One approach~\cite{Catterall:2007kn, Catterall:2014vka, Schaich:2014pda, Catterall:2015ira} combines the gauge and scalar fields into a complexified gauge field, leading to $\U{N} = \SU{N} \otimes \Uone$ gauge invariance and non-compact lattice gauge links $\left\{\cU, \cUbar\right\}$ with a flat measure.
The fermion fields are twisted in the same way as the supercharges, obtaining the same lattice gauge transformations as the bosonic degrees of freedom.
Although the U(1) sector decouples in the continuum, at non-zero lattice spacing it can introduce unwanted artifacts at strong coupling, and ongoing work is searching for good ways to suppress these~\cite{Catterall:2015ira, Catterall:2017lub, Catterall:2018laz}.
For $d = 2$ a different approach~\cite{Sugino:2003yb, Sugino:2004qd, Kadoh:2009rw, Hanada:2010kt, Hanada:2011qx, Matsuura:2014kha, Hanada:2017gqc} works with compact gauge links and gauge group SU($N$), at the cost of imposing an admissibility condition to resolve a huge degeneracy of vacua (but see \refcite{Matsuura:2014pua}), which becomes more problematic in higher dimensions~\cite{Sugino:2004uv, Catterall:2009it}. 
This formulation has been used by several numerical studies of the $Q = 4$~\cite{Hanada:2009hq, Suzuki:2007jt, Kanamori:2007ye, Kanamori:2007yx, Kanamori:2008bk, Kanamori:2008yy, Kanamori:2009dk, Hanada:2010qg, Kamata:2016xmu} and $Q = 16$~\cite{Giguere:2015cga, Kadoh:2017mcj} theories. 

The right plot of \fig{fig:phase_diagrams} shows recent results from \refcite{Catterall:2017lub} for the phase diagram of two-dimensional $Q = 16$ SYM, using the non-compact twisted construction described above~\cite{Catterall:2008dv, Catterall:2010fx, Catterall:2011aa, Jha:2018acq}. 
The system is formulated on an $r_L \X r_{\be}$ torus, with $r_{\be} = 1 / T$ the inverse dimensionless temperature while $r_L = L_{\text{dim}} \sqrt{\la_{\text{dim}}}$ is the corresponding dimensionless length of the spatial cycle.
At high temperatures (small $r_{\be}$), the fermions pick up a large thermal mass and the system reduces to a one-dimensional bosonic QM.
In this limit (at large $N$), Refs.~\cite{Aharony:2004ig, Kawahara:2007fn, Mandal:2009vz, Azuma:2014cfa} predict a `spatial deconfinement' transition as $r_L$ decreases, signalled by a non-zero spatial Wilson line $\Tr{\prod_{x_i} \cU_x(x_i, t)}$.
It is currently unclear whether this is a single first-order transition~\cite{Azuma:2014cfa} or two nearby second- and third-order transitions~\cite{Kawahara:2007fn, Mandal:2009vz}.

In the low-temperature (large-$r_{\be}$) limit, there is a large-$N$ holographic prediction for a similar transition.
Here the large-$r_L$ spatially confined phase is conjectured to be dual to a homogeneous black string with a horizon wrapping around the spatial cycle, while the small-$r_L$ spatially deconfined phase corresponds to a localized black hole.
As in the BMN case, the holographic analyses require challenging numerical SUGRA constructions~\cite{Dias:2017uyv} of these dual black hole and black string geometries.
The lattice results for the spatial deconfinement transition (with $N = 12$ and fixed lattice sizes $32\X 4$, $24\X 4$, $24\X 6$, $24\X 9$, $24\X 12$ and $18\X 12$) reproduce the high-temperature bosonic QM expectations quite well and are consistent with holography at lower temperatures, albeit with rapidly increasing uncertainties.
At low temperatures a scalar potential is added to the lattice action and then extrapolated to zero in order to avoid the thermal instability mentioned above for SYM QM.
\refcite{Catterall:2017lub} also calculates the internal energies of the dual black hole and black string, in both phases finding consistency with holographic expectations within large uncertainties.
It will be interesting to see future work improve upon these results, ideally accessing lower temperatures in addition to gaining control over extrapolations to the large-$N$ continuum limit.

Two-dimensional SYM also possesses rich zero-temperature dynamics that are important to explore non-perturbatively, in addition to studying the thermal behavior discussed above.
For example, Refs.~\cite{Witten:1995im, Fukaya:2006mg} argue that the `meson' spectrum of the $Q = 4$ theory should include a massless supermultiplet, unlike the $d = 4$ $\cN = 1$ SYM of which this is the dimensional reduction.
A recent lattice calculation using straightforward Wilson fermions observes such a massless multiplet~\cite{August:2018esp}, and also checks for spontaneous supersymmetry breaking, which \refcite{Hori:2006dk} suggests might occur for this theory.
No evidence of spontaneous supersymmetry breaking is seen, consistent with another recent lattice study~\cite{Catterall:2017xox} and older work~\cite{Kanamori:2007ye, Kanamori:2007yx, Kanamori:2009dk} using twisted formulations. 

\section{\label{sec:min}Minimally supersymmetric Yang--Mills ($\cN = 1$ SYM) in four dimensions} 
Returning to four dimensions, we can note that most of the supersymmetry-violating operators discussed in \secref{sec:intro} involve scalar fields, viz.\ the scalar mass terms, Yukawa couplings, and quartic operators.
This implies a vast reduction of fine-tuning for $\cN = 1$ SYM, the only $d = 4$ supersymmetric gauge theory with no scalar fields.
This theory consists of a SU($N$) gauge field and its superpartner gaugino, a massless Majorana fermion transforming in the adjoint representation of SU($N$).
The only relevant (or marginal) operator that may need to be fine-tuned to obtain the correct continuum limit is the gaugino mass~\cite{Curci:1986sm, Suzuki:2012pc}.
We can even avoid this single fine-tuning by working with Ginsparg--Wilson (overlap or domain-wall) lattice fermions that preserve chiral symmetry and protect the gaugino mass against large additive renormalization.
Although the axial anomaly breaks the classical U(1) $R$-symmetry of $\cN = 1$ SYM to its $Z_{2N}$ subgroup, this discrete global symmetry suffices to forbid a gaugino mass.
Gaugino condensation, $\vev{\la\la} \ne 0$, spontaneously breaks $Z_{2N} \to Z_2$.

However, in large part due to their computational expense, there have been no Ginsparg--Wilson studies of $\cN = 1$ SYM for most of the past decade~\cite{Giedt:2008xm, Endres:2009yp, Kim:2011fw}.
Instead, current work uses improved Wilson fermions and fine-tunes the gaugino mass to recover both chiral symmetry and supersymmetry in the continuum limit.
One major effort by the DESY--M{\"u}nster--Regensburg--Jena Collaboration, currently using clover improvement, has made significant progress in recent years~\cite{Bergner:2014saa, Bergner:2014dua, Bergner:2014ska, Bergner:2015adz, Ali:2017iof, Ali:2018dnd, Ali:2018fbq, Bergner:2018unx}. 
A second group recently began exploring a SYM analogue of the twisted-mass fermion action~\cite{Steinhauser:2017xqc, Steinhauser:2018wep}, aiming to improve the formation of composite supermultiplets at non-zero gaugino masses and lattice spacings, and thereby gain better control over the chiral and continuum extrapolations.

\begin{figure}[bp]
  \centering
  \includegraphics[height=0.36\linewidth]{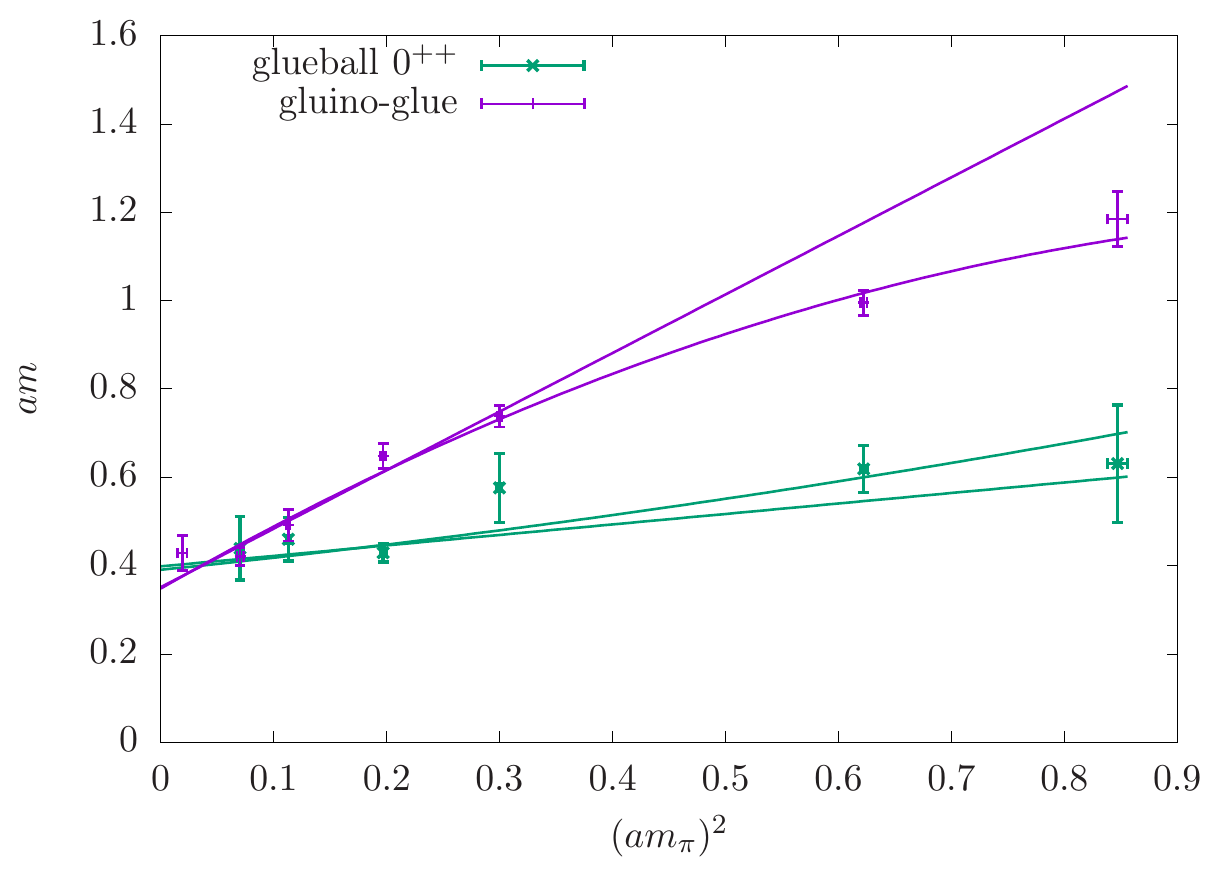}\hfill \includegraphics[height=0.36\linewidth]{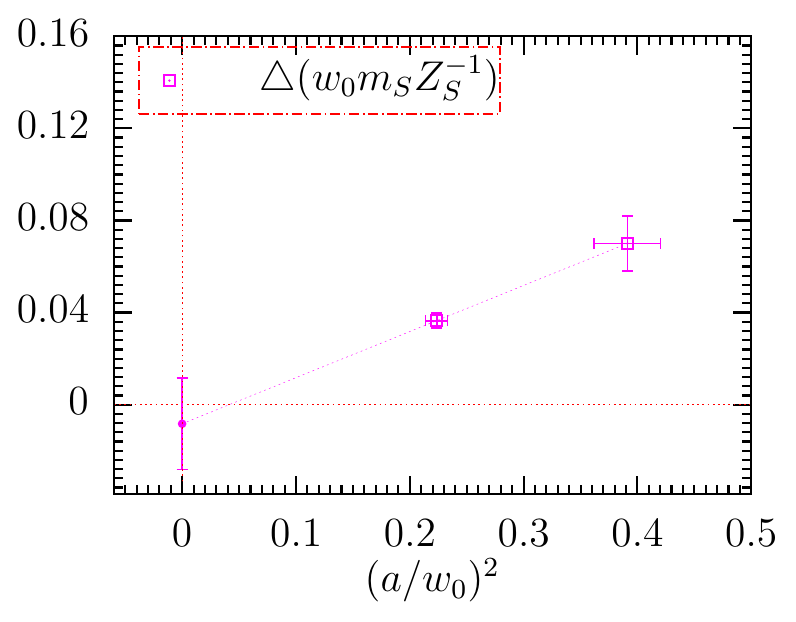}
  \caption{\label{fig:N1clover}Recent results from lattice $\cN = 1$ SYM calculations using gauge group SU(3) and Wilson-clover fermions.  \textbf{Left:} $0^{++}$ `glueball' and fermionic `gluino--glue' particle masses vs.\ the `adjoint pion' mass squared, from \protect\refcite{Ali:2018dnd}.  The $m_{\pi}^2 \to 0$ extrapolations of these masses agree within uncertainties even at a fixed lattice spacing, supporting the formation of supermultiplets expected in the chiral continuum limit.  \textbf{Right:} A measure of supersymmetry-breaking discretization artifacts (defined in the text) is consistent with vanishing $\propto$$a^2$ in the $a \to 0$ continuum limit, from \protect\refcite{Ali:2018fbq}.}
\end{figure}

The larger number of dimensions requires considering much smaller $N \ll 10$ compared to the lower-dimensional work discussed above, to keep computational costs under control.
Current efforts study only gauge groups SU(2)~\cite{Bergner:2014saa, Bergner:2014dua, Bergner:2014ska, Bergner:2015adz, Ali:2017iof} and SU(3)~\cite{Ali:2018dnd, Ali:2018fbq, Bergner:2018unx, Steinhauser:2017xqc, Steinhauser:2018wep, Ali:2018rln}. 
The left plot of \fig{fig:N1clover} shows recent SU(3) results from \refcite{Ali:2018dnd} for the masses of two composite states expected to form (part of) a degenerate multiplet in the supersymmetric continuum chiral limit~\cite{Veneziano:1982ah, Farrar:1997fn}: the $0^{++}$ `glueball' and the fermionic `gluino--glue' particle.
Even at a fixed lattice spacing the chiral extrapolations of these masses agree within uncertainties.
These signs of supermultiplet formation appear much clearer compared to earlier SU(2) results~\cite{Bergner:2015adz}, presumably due to either or both the larger $N$ and the use of clover improvement instead of stout smearing.

The chiral extrapolations in \fig{fig:N1clover} are carried out by computing the mass of an `adjoint pion' defined in partially quenched chiral perturbation theory~\cite{Munster:2014cja} and taking the limit $m_{\pi}^2 \to 0$.
While two-point functions for the physical composite states of $\cN = 1$ SYM all involve fermion-line-disconnected diagrams, $m_{\pi}$ is measured from just the connected part of the correlator for the $\eta'$-like `gluinoball'. 
Supersymmetric Ward identities provide an alternative means to determine the critical $\ka_c$ corresponding to the chiral limit. 
The difference between these two determinations of $\ka_c$ can be considered a measure of the supersymmetry-breaking discretization artifacts, which is shown for two lattice spacings in right plot of \fig{fig:N1clover}.
The two available points are consistent with the artifacts vanishing $\propto$$a^2$ as expected for clover fermions, supporting the restoration of supersymmetry in the chiral continuum limit.

As for QCD, many other lattice $\cN = 1$ SYM investigations may be carried out in addition to calculations of the spectrum, Ward identities, and the gaugino condensate $\vev{\la\la}$~\cite{Giedt:2008xm, Endres:2009yp, Kim:2011fw}.
These include explorations of the finite-temperature phase diagram, with Refs.~\cite{Bergner:2014saa, Bergner:2018qpb} reporting that deconfinement (spontaneous center symmetry breaking) and chiral symmetry restoration appear to occur at the same temperature, which was not known a priori.
Refs.~\cite{Bergner:2014dua, Bergner:2018unx} investigate the phase diagram on $\Rbb^3 \X S^1$ with a small radius for the compactified temporal direction.
Comparing thermal and periodic boundary conditions (BCs) for the gauginos, they find evidence that periodic BCs allow the confined, chirally broken phase to persist for weak couplings where analytic semi-classical methods~\cite{Poppitz:2012sw} may be reliable.
In addition, there is ongoing work to construct a SYM gradient flow that is consistent with supersymmetry in Wess--Zumino gauge~\cite{Kadoh:2018qwg}, which could be used to define a renormalized supercurrent and help guide fine-tuning~\cite{Hieda:2017sqq, Kasai:2018koz}. 
The ordinary non-supersymmetric gradient flow is already used by many lattice $\cN = 1$ SYM projects, to set the scale (as in the right plot of \fig{fig:N1clover}) and improve signals for observables such as the gaugino condensate~\cite{Bergner:2018qpb}.
Finally, given the progress in algorithms and computing hardware over the past decade, it seems worthwhile to revisit calculations with Ginsparg--Wilson fermions, which could complement and check the ongoing Wilson-fermion work.

\section{\label{sec:max}Maximally supersymmetric Yang--Mills ($\cN = 4$ SYM) in four dimensions} 
In \secref{sec:lowd} we discussed why the twisted (and orbifolded) constructions of SYM with exact supersymmetry at non-zero lattice spacing require $Q \geq 2^d$ supercharges.
In $d = 4$ dimensions, this constraint picks out another special case, $\cN = 4$ SYM with $Q = 16$, for which a single `twisted-scalar' supercharge \cQ is preserved.
This theory consists of a SU($N$) gauge field, four Majorana fermions and six real scalars, all massless and transforming in the adjoint representation of SU($N$) as usual.
Thanks to its many supersymmetries, large SU(4)$_R$ symmetry and conformal symmetry, $\cN = 4$ SYM is widely studied throughout theoretical physics (especially in its large-$N$ planar limit).
Among many other important roles, it is the conformal field theory of the original AdS/CFT holographic duality~\cite{Maldacena:1997re}, and provided early insight into S-duality~\cite{Osborn:1979tq}.
Lattice field theory in principle enables non-perturbative investigations of this theory even away from the planar regime.

\begin{figure}[bp]
  \centering
  \includegraphics[width=0.48\linewidth]{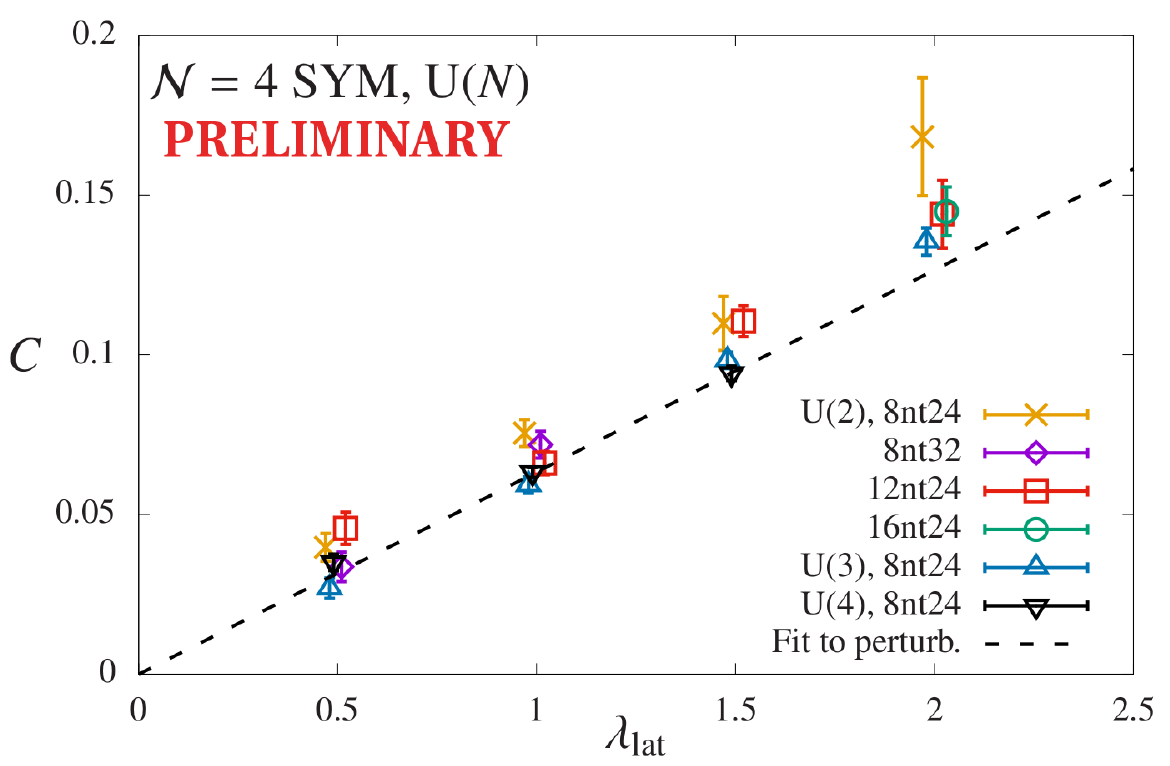}\hfill \includegraphics[width=0.48\linewidth]{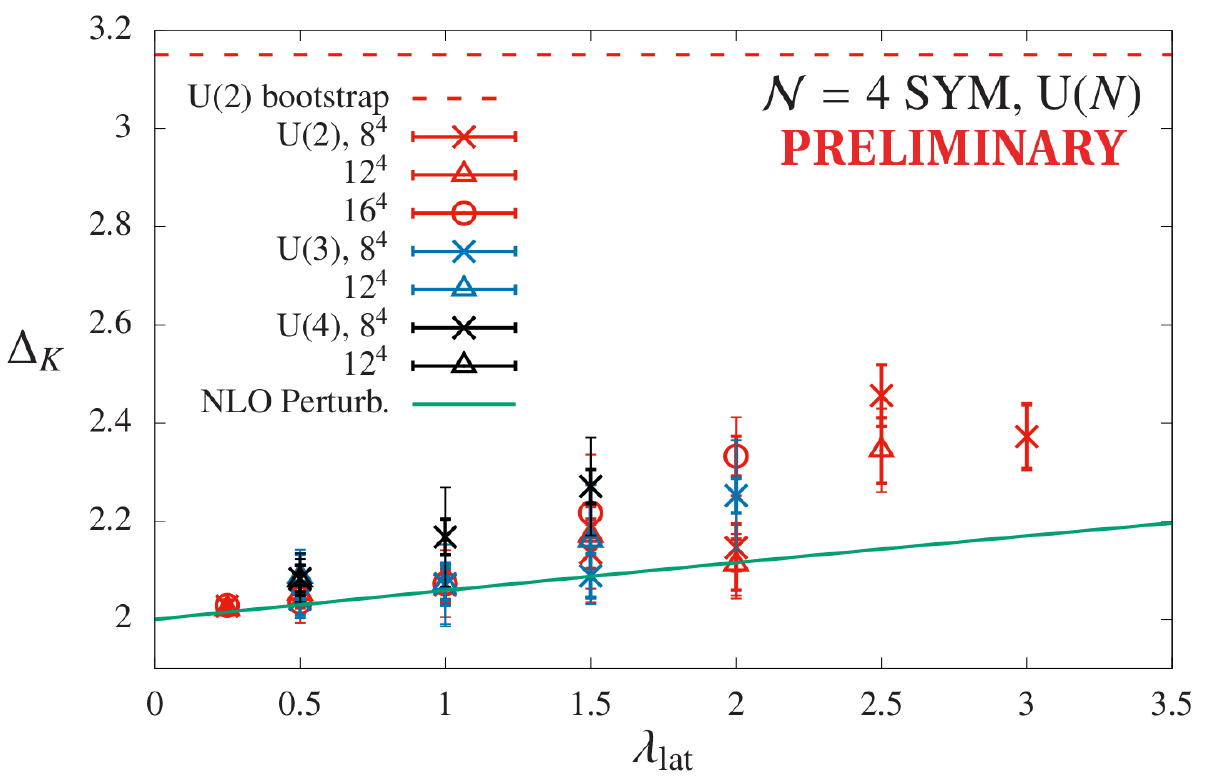}
  \caption{\label{fig:N4SYM}Preliminary results from ongoing four-dimensional lattice $\cN = 4$ SYM calculations with gauge groups U(2), U(3) and U(4).  \textbf{Left:} The static potential Coulomb coefficient, from $L^3\X N_t$ lattices with $L \leq 16$ and $N_t \leq 32$, appears consistent with leading-order perturbation theory (black dashed line) for $\lalat \leq 2$.  \textbf{Right:} The Konishi scaling dimension, from MCRG stability matrix analyses of $L^4$ lattices with $L \leq 16$, also appears consistent with perturbation theory (and well below bootstrap bounds) for $\lalat \lsim 3$.}
\end{figure}

On the lattice, the bosonic fields are combined into five-component complexified gauge links $\left\{\cU, \cUbar\right\}$, implying the $A_4^*$ lattice structure of five basis vectors symmetrically spanning four dimensions~\cite{Catterall:2009it, Kaplan:2005ta, Unsal:2006qp, Catterall:2007kn, Damgaard:2008pa}. 
A single fine-tuning of a marginal operator may be required to recover the continuum twisted rotation symmetry from the $S_5$ point-group symmetry of the $A_4^*$ lattice, which in turn restores the 15 supersymmetries broken by the lattice discretization~\cite{Catterall:2014mha, Catterall:2013roa, Catterall:2014vga}. 
Most numerical calculations so far fix the corresponding coefficient to its classical value.
These calculations also have to regulate flat directions in both the SU($N$) and U(1) sectors.
A simple (soft $\cQ$-breaking) scalar potential suffices to lift the SU($N$) flat directions, and is removed in continuum extrapolations.
The U(1) sector is more challenging, and ongoing work is searching for good ways to handle it~\cite{Catterall:2015ira, Catterall:2017lub, Catterall:2018laz}. 
The results shown in \fig{fig:N4SYM} lift the U(1) flat directions by modifying the moduli equations in a way that preserves the $\cQ$ supersymmetry.
At least for 't~Hooft couplings $\lalat \leq 2$ this results in effective $\cO(a)$ improvement indicated by \cQ Ward identity violations vanishing $\propto$$a^2$ in the continuum limit~\cite{Catterall:2015ira}.
The resulting lattice action is rather complicated, motivating the public development of high-performance parallel code~\cite{Schaich:2014pda} for lattice $\cN = 4$ SYM and lower-dimensional SYM theories at \texttt{\href{https://github.com/daschaich/susy/}{github.com/daschaich/susy}}.

Figure~\ref{fig:N4SYM} presents some preliminary results from ongoing lattice $\cN = 4$ SYM calculations.
The left plot considers the static potential $V(r)$, which is found to be coulombic at all accessible 't~Hooft couplings~\cite{Catterall:2014vka, Catterall:2012yq, Schaich:2016jus}, as expected. 
Fitting (tree-level-improved~\cite{Schaich:2016jus}) lattice data to the Coulomb potential $V(r) = A - C / r$ predicts the Coulomb coefficient $C(\la)$ shown in the figure.
There is a famous holographic prediction~\cite{Rey:1998ik, Maldacena:1998im} that in the regime $N \to \infty$ and $\la \to \infty$ with $\la \ll N$ this quantity should behave as $C(\la) \propto \sqrt{\la}$ up to $\cO\left(\frac{1}{\sqrt{\la}}\right)$ corrections, and more general analytic results have been obtained in the $N = \infty$ planar limit~\cite{Gromov:2016rrp}.
The lattice results for $N \leq 4$ and $\lalat \leq 2$ do not show such behavior and instead look consistent with leading-order perturbation theory.
The dashed black line is a fit of the U(4) data to the leading perturbative expression $C(\la) = b\lalat / (4\pi)$, where the fit parameter $b = 0.795(13)$ converts the input lattice 't~Hooft coupling to the expected continuum normalization.
Higher-order perturbative corrections for $C(\la)$ are suppressed by powers of $\frac{\la}{2\pi^2}$~\cite{Pineda:2007kz, Stahlhofen:2012zx, Prausa:2013qva}, suggesting that this apparent leading-order behavior for $\lalat \leq 2$ should not be surprising.

The right plot of \fig{fig:N4SYM} considers the scaling dimension $\De_K(\la) = 2 + \ga_K(\la)$ of the simplest conformal primary operator of $\cN = 4$ SYM, the Konishi operator $\cO_K = \sum_I \Tr{X^I X^I}$, where $X^I$ are the scalar fields (obtained from a polar decomposition of the complexified lattice gauge links).
There are again both perturbative~\cite{Fiamberti:2008sh, Bajnok:2008bm, Velizhanin:2008jd} and holographic~\cite{Gubser:1998bc, Gromov:2009zb} predictions for $\De_K$.
The former are also relevant for the strong-coupling regime $\la \gg N$~\cite{Beem:2013hha}, due to the conjectured S-duality of the theory, which relates its spectrum of anomalous dimensions under the interchange $\frac{4\pi N}{\la} \llra \frac{\la}{4\pi N}$.
In addition, the superconformal bootstrap program has obtained bounds on the maximum value $\ga_K$ can reach across all $\la$~\cite{Beem:2013qxa, Beem:2016wfs}.
The lattice results in this figure for $\lalat \lsim 3$ again appear consistent with perturbation theory.
They are obtained from Monte Carlo renormalization group (MCRG) stability matrix analyses~\cite{Swendsen:1979gn}, with systematic uncertainties estimated by varying the number of interpolating operators in the stability matrix (with different operators obtained by using different amounts of smearing).
Additional systematic uncertainties still to be quantified include sensitivity to the lattice volume and the number of RG blocking steps.
The stability matrix also includes the related `SUGRA' or $20'$ operator $\cO_S^{IJ} = \Tr{X^{\{I} X^{J\}}}$, whose scaling dimension is fixed to its protected value $\De_S = 2$. 

Existing numerical calculations only scratch the surface of the investigations that could in principle be pursued by lattice $\cN = 4$ SYM.
One important task is to push existing studies like those in \fig{fig:N4SYM} to stronger 't~Hooft couplings, in order to make contact with holographic predictions and ideally investigate the behavior of the system around the S-dual point $\la_{\text{sd}} = 4\pi N$.
The discussion of sign problems in the next section suggests that this is likely to be challenging.
An alternative possibility is to study S-duality at currently accessible couplings by adjusting the scalar potential so that the system moves onto the Coulomb branch of the moduli space where its U($N$) gauge invariance is higgsed to U(1)$^N$.
In this context S-duality relates the masses of the U(1)-charged elementary `$W$~bosons' and the magnetically charged topological 't~Hooft--Polyakov monopoles~\cite{Osborn:1979tq}, each of which may be accessible from lattice calculations with either C-periodic or twisted BCs~\cite{Giedt:2016thz}.
The finite-temperature behavior of lattice $\cN = 4$ SYM will also be interesting to explore.
In particular, there is motivation~\cite{Hanada:2016jok} to study the free energy, for which the weak-coupling perturbative prediction~\cite{Fotopoulos:1998es} and the holographic strong-coupling calculation~\cite{Gubser:1998bc} differ by a factor of $\frac{3}{4}$.

\section{\label{sec:future}Challenges for the future} 
Although the recent progress of lattice supersymmetry is substantial, it is largely concentrated in the three areas discussed above where significant simplifications are possible.
Within those three areas we have already considered several compelling directions for future work, ranging from improved control over large-$N$ continuum extrapolations in lower dimensions, to revisiting $\cN = 1$ SYM with Ginsparg--Wilson fermions, and reaching stronger 't~Hooft couplings in $\cN = 4$ SYM calculations.
In addition, it will be important for efforts to expand beyond these domains and tackle more challenging subjects where such simplifications do not appear to be available.
We conclude this brief review by touching on some of these subjects, highlighting SQCD and the possibility of sign problems in supersymmetric lattice systems.

\paragraph{Supersymmetric QCD:} 
Adding matter multiplets (`quarks' and `squarks' not necessarily in the fundamental representation) to the four-dimensional lattice $\cN = 1$ SYM work discussed in \secref{sec:min} would enable investigations of many important phenomena, including (metastable) dynamical supersymmetry breaking, conjectured electric--magnetic dualities and RG flows to known conformal IR fixed points. 
The downside is that many more supersymmetry-violating operators appear, and the fine-tuning challenge becomes enormously harder.
Even exploiting the continuum-like flavor symmetries offered by Ginsparg--Wilson fermions, \refcite{Giedt:2009yd} counts $\cO(10)$ operators to be fine-tuned, depending on the gauge group and matter content.
In this context working with Ginsparg--Wilson fermions appears to be especially strongly motivated, with Refs.~\cite{Giedt:2009yd, Elliott:2008jp} arguing that this may allow most or all of the scalar masses, Yukawas and quartic couplings to be fine-tuned ``offline'' through multicanonical reweighting, which could vastly reduce computational costs.

That said, as in the case of $\cN = 1$ SYM, work currently underway uses Wilson fermions and has to face the full fine-tuning head-on.
One tactic for approaching this challenge is to use lattice perturbation theory to guide numerical calculations~\cite{Costa:2017rht, Costa:2018mvb, Wellegehausen:2018opt}.
Another is to omit the scalar fields at first, and warm up by studying the gauge--fermion theory including both (adjoint) gauginos and (fundamental) quarks~\cite{Bergner:2018znw}, which also provides connections to composite Higgs investigations that are reviewed by another contribution to these proceedings~\cite{Witzel:2019jbe}.
These four-dimensional efforts are just getting underway.

Following the logic of \secref{sec:lowd}, it may prove advantageous to first investigate simpler systems in fewer than four dimensions.
In 0+1~dimensions, for example, Refs.~\cite{Filev:2015cmz, Asano:2016xsf, Asano:2016kxo} consider the Berkooz--Douglas matrix model~\cite{Berkooz:1996is}, which adds $N_f$ fundamental multiplets to $Q = 16$ SYM~QM (preserving half of the supercharges in the continuum).
As for the case of two- and three-dimensional SYM, more effort has been devoted to constructing clever lattice formulations of $d = 2$ and $d = 3$ SQCD~\cite{Matsuura:2008cfa, Sugino:2008yp, Kikukawa:2008xw, Kadoh:2009yf, Joseph:2013jya, Joseph:2013bra, Joseph:2014bwa} compared to carrying out numerical calculations~\cite{Catterall:2015tta}. 

That one numerical calculation~\cite{Catterall:2015tta} uses a generalization of the twisted formulation to realize a quiver construction of two-dimensional $Q = 4$ SQCD that still preserves one of the supercharges at non-zero lattice spacing~\cite{Matsuura:2008cfa, Sugino:2008yp}.
The starting point is three-dimensional 8-supercharge SYM on a lattice with only two slices in the third direction.
The twisted formulation can be generalized to have different gauge groups U($N$) and U($F$) on each slice, with the bosonic and fermionic fields that connect the two slices transforming in the bifundamental representation of $\U{N}\X \U{F}$.
Decoupling the U($F$) slice then leaves behind a two-dimensional U($N$) theory with half the supercharges ($Q = 4$) and $F$ massless fundamental matter multiplets.
This same procedure works for $Q = 8$ SQCD in two and three dimensions~\cite{Joseph:2013jya, Joseph:2013bra}, and may be generalizable to higher representations~\cite{Joseph:2014bwa}.
\refcite{Catterall:2015tta} compares U(2) SQCD with $F = 3$ vs.\ U(3) SQCD with $F = 2$, observing dynamical supersymmetry breaking for $N > F$ and confirming that the resulting goldstino is consistent with masslessness in the infinite-volume limit.\footnote{Lower-dimensional Wess--Zumino models could also be a useful setting for future lattice studies of dynamical supersymmetry breaking and goldstinos~\cite{Wozar:2011gu, Steinhauer:2014yaa, Baumgartner:2014nka, Baumgartner:2015qba, Baumgartner:2015zna}.}

\paragraph{Sign problems:} 
Another challenge is that some of the supersymmetric lattice systems discussed above may suffer from a sign problem, at least in certain regimes.
Since the gauginos are Majorana fermions, integrating over them produces the pfaffian of the fermion operator, which can fluctuate in sign even when the determinant would be positive.
Writing a generic complex pfaffian as ${\pf \cD = |\pf \cD| e^{i\al}}$, only its magnitude is included in the `phase-quenched' RHMC studies presented above.
The phase-quenched observables $\vev{\cO}_{\text{pq}}$ need to be reweighted, $\vev{\cO} = \vev{\cO}_{\text{pq}} / \vev{e^{i\al}}_{\text{pq}}$, with a sign problem appearing when $\vev{e^{i\al}}_{\text{pq}} = Z / Z_{\text{pq}}$ vanishes within statistical uncertainties.
(See \refcite{deForcrand:2010ys} for a brief introduction to sign problems.)
In particular, in lattice calculations with periodic BCs for all fields, the partition function $Z$ is the Witten index and must vanish for any theory that can exhibit spontaneous supersymmetry breaking~\cite{Witten:1982df}, implying a severe sign problem.

For Wilson-fermion $\cN = 1$ SYM the pfaffian is real and its sign can be computed efficiently~\cite{Bergner:2011zp}.
Recent clover calculations report $\vev{e^{i\al}}_{\text{pq}} \approx 1$, with the situation improving further as the lattice spacing decreases~\cite{Ali:2018dnd}.
However, $\vev{e^{i\al}}_{\text{pq}}$ is expected to decrease exponentially in the lattice volume, and the situation is likely to be worse for SQCD. 
Directly evaluating the pfaffian is much more computationally expensive, and has been done mostly for SYM QM and $d = 2$ SYM, where sign problems also appear to be well under control~\cite{Catterall:2009xn, Filev:2015hia, Hanada:2010qg, Kamata:2016xmu, Catterall:2011aa, August:2018esp, Catterall:2017xox}. 

\begin{figure}[bp]
  \centering
  \includegraphics[width=0.48\linewidth]{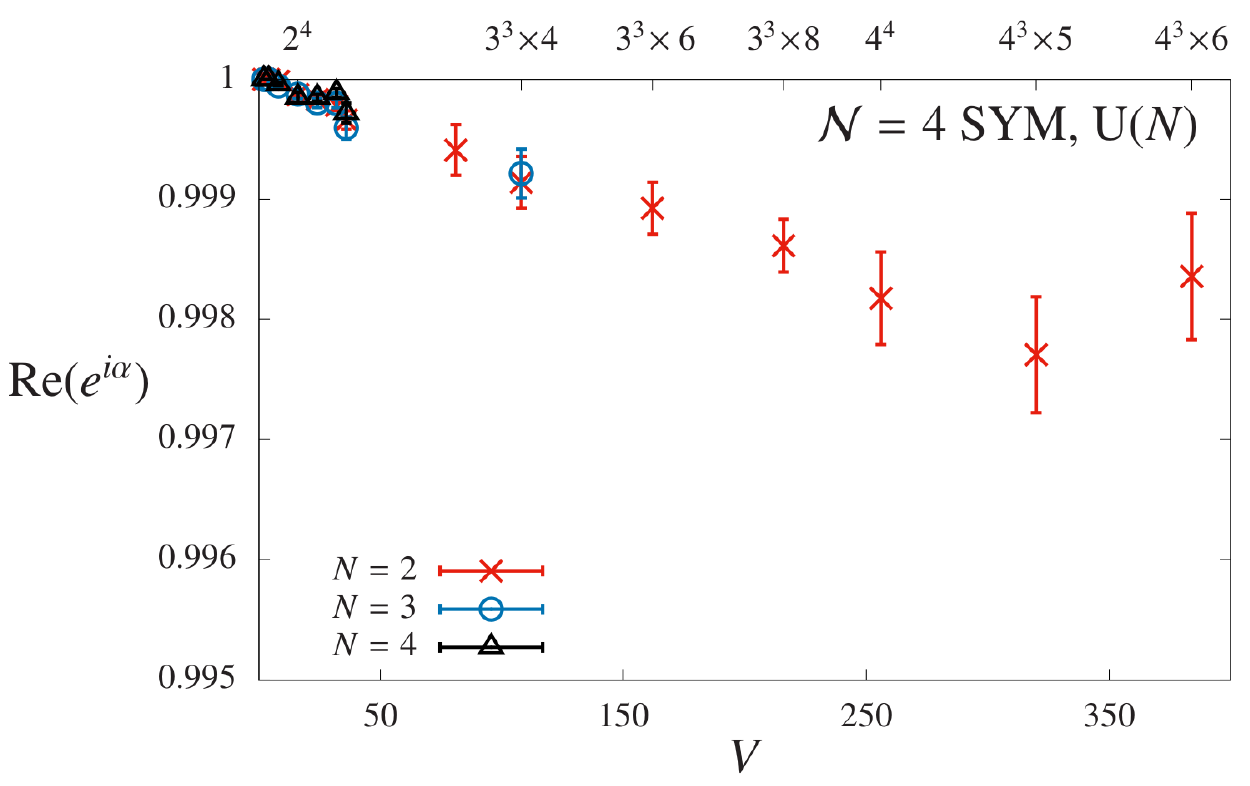}\hfill \includegraphics[width=0.48\linewidth]{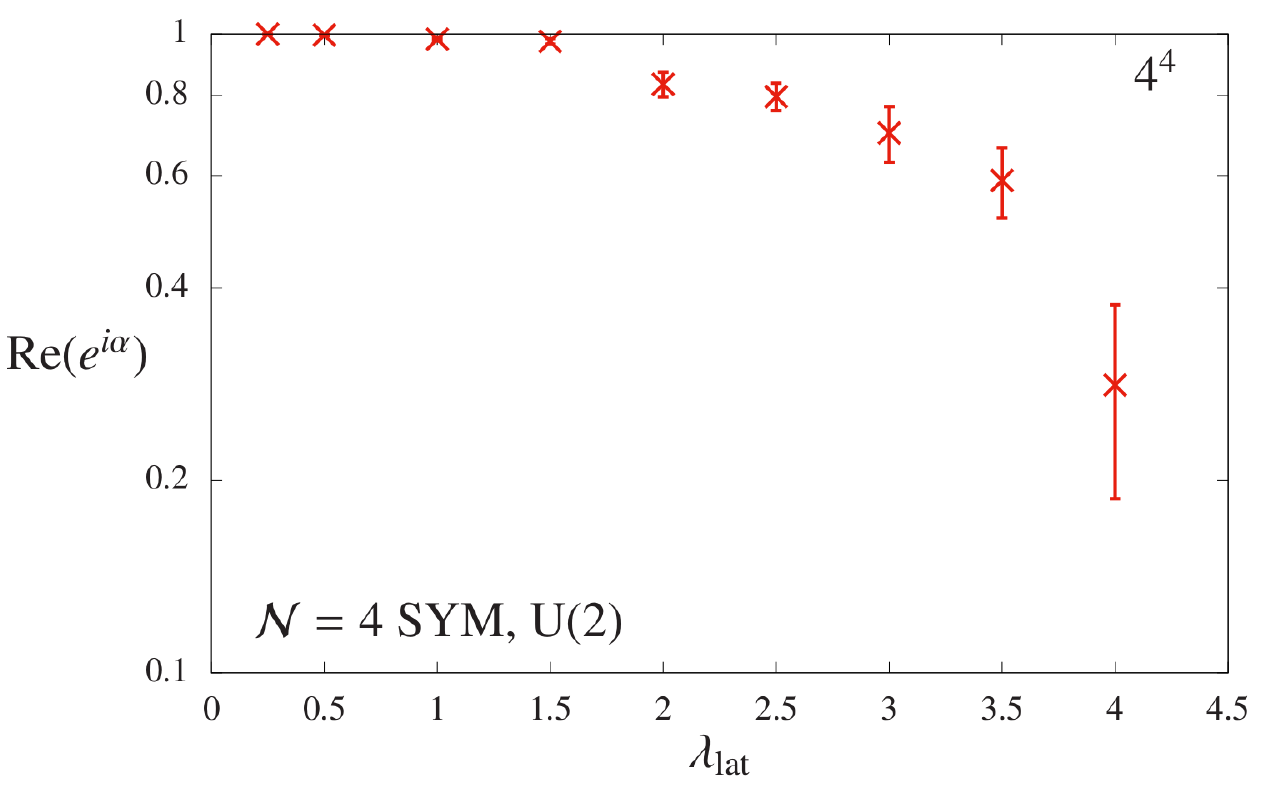}
  \caption{\label{fig:N4phase}Results for the phase of the pfaffian $\vev{\mbox{Re} \left(e^{i\al}\right)}_{\text{pq}} \approx \vev{e^{i\al}}_{\text{pq}}$ from lattice $\cN = 4$ SYM in four dimensions.  \textbf{Left:} With fixed 't~Hooft coupling $\lalat = 0.5$, only per-mille-level fluctuations are observed for U($N$) gauge groups with $N = 2$, 3 and 4, up to the largest accessible volumes.  Adapted from \protect\refcite{Catterall:2014vga}.  \textbf{Right:} On a fixed $4^4$ lattice volume, the phase fluctuations increase significantly for stronger couplings $\lalat \gsim 2$, obstructing studies of $\lalat \gsim 4$ with this lattice action. Adapted from \protect\refcite{Schaich:2015daa}.}
\end{figure}

Figure~\ref{fig:N4phase} presents results for the pfaffian phase of lattice $\cN = 4$ SYM in four dimensions, adapted from Refs.~\cite{Catterall:2014vga, Schaich:2015daa}, where only small $N$ and small lattice volumes are computationally accessible.
(Each pfaffian measurement for a single $4^4$ lattice with $N = 2$ takes approximately 50~hours on 16~cores, and costs scale with the cube of the number of fermion degrees of freedom~\cite{Schaich:2014pda}.)
In the left plot, only small per-mille-level phase fluctuations are observed on all accessible volumes with fixed 't~Hooft coupling $\lalat = 0.5$.
In particular, the expected exponential suppression of $\vev{e^{i\al}}_{\text{pq}}$ with the lattice volume is not visible; instead the largest volumes for gauge group U(2) produce results that are constant within uncertainties.
In the right plot, however, we see phase fluctuations increasing significantly for stronger 't~Hooft couplings $\lalat \gsim 2$.
This appears to be one of the main obstacles to reaching the stronger couplings of interest in order to directly probe holography and S-duality, with calculations using this lattice action largely limited to $\lalat \lsim 4$.

\paragraph{Final remarks:} 
Non-perturbative lattice investigations of supersymmetric QFTs are important and challenging, making this a field in which we can expect to see a great deal more work in the future.
It is encouraging that there has been so much recent progress in lattice studies of four-dimensional $\cN = 1$ SYM and $\cN = 4$ SYM, along with their dimensional reductions to $d < 4$.
This brief overview has also omitted coverage of advances in other areas, including theories without gauge invariance such as Wess--Zumino models and sigma models~\cite{Wozar:2011gu, Steinhauer:2014yaa, Baumgartner:2014nka, Baumgartner:2015qba, Baumgartner:2015zna, Aoki:2017iwi, Kadoh:2018hqq, Kadoh:2018ele, Kadoh:2018ivg}, the lattice regularization of the Green--Schwarz superstring worldsheet sigma model~\cite{McKeown:2013vpa, Bianchi:2016cyv, Forini:2017ene}, and proposals for lattice formulations of a mass-deformed $\cN = 2^*$ SYM theory with $Q = 8$ in four dimensions~\cite{Joseph:2017nap} and of $Q = 16$ SYM in five dimensions~\cite{Joseph:2016tlc}.
While there are clear challenges that will be difficult to overcome, in particular concerning supersymmetric QCD and sign problems, overall the prospects of lattice supersymmetry are bright, with many compelling directions for future investigations.

\vspace{20 pt}
\noindent \textsc{Acknowledgments:}~I thank the organizers of Lattice 2018 for the invitation to present this overview, and for all their work to manage the conference.
My participation in the conference was supported by a travel grant from the Faculty of Science at the University of Bern.
Enrico Rinaldi, Jun Nishimura, Hiroto So, Marios Costa, Georg Bergner and Bj\"orn Wellegehausen provided helpful information about their recent work.
I have benefited from collaboration on lattice supersymmetry with Simon Catterall, Poul Damgaard, Tom DeGrand, Joel Giedt, Raghav Jha, Anosh Joseph and Toby Wiseman.

\clearpage
\bibliographystyle{utphys}
\bibliography{lattice18}
\end{document}